\documentclass[a4paper, 11pt]{article}

\usepackage{amsmath,amsthm,amssymb,graphicx,url,verbatim,bbm,enumitem,multirow,array,diagbox}
\usepackage{tikz}
 \usepackage[margin=2cm]{geometry}
\usetikzlibrary{automata}
\usepackage{bm}
\DeclareFontFamily{U}{matha}{}
\DeclareFontShape{U}{matha}{m}{n}{
	<-5.5>    matha5
	<5.5-6.5> matha6 
	<6.5-7.5> matha7
	<7.5-8.5> matha8
	<8.5-9.5> matha9
	<9.5-11>  matha10
	<11->     matha12
}{}
\DeclareSymbolFont{matha}{U}{matha}{m}{n}
\DeclareFontSubstitution{U}{matha}{m}{n}
\DeclareFontFamily{U}{mathx}{\hyphenchar\font45}
\DeclareFontShape{U}{mathx}{m}{n}{<-> mathx10}{}
\DeclareSymbolFont{mathx}{U}{mathx}{m}{n}
\DeclareFontSubstitution{U}{mathx}{m}{n}

\DeclareMathDelimiter{\ldbrack}{4}{matha}{"76}{mathx}{"30}
\DeclareMathDelimiter{\rdbrack}{5}{matha}{"77}{mathx}{"38}

\numberwithin{equation}{section}

\theoremstyle{plain}
 \newtheorem{theorem}	[equation]	{Theorem}
\newtheorem{conjecture}	[equation]	{Conjecture}
 \newtheorem{corollary}	[equation]	{Corollary}
 
 \newtheorem{example}	[equation]	{Example}
 \newtheorem{lemma}		[equation]	{Lemma}

 \newtheorem{proposition}[equation]	{Proposition}
 
\theoremstyle{definition}
 \newtheorem{claim}		[equation]	{Claim}
\newtheorem*{claim*}{Claim}

\newcommand{\neighbourhood}{N}

\newcommand{\inn}[1]{#1_\mathrm{in}}
\newcommand{\out}[1]{#1_\mathrm{out}}

\newcommand{\NIn}{\inn{\neighbourhood}}

\newcommand{\NOut}{\out{\neighbourhood}}

\newcommand{\IG}{\mathbb{D}}

\newcommand{\functions}{\mathrm{F}}

\newcommand{\matrices}{\mathrm{M}}

\newcommand{\dmin}{d_{\min}}

\newcommand{\dH}{d_\mathrm{H}}

\newcommand{\GF}{\mathrm{GF}}

\newcommand{\rank}{\mathrm{rk}}

\renewcommand{\)}{\rdbrack}

\newcommand{\Z}{\mathbb{Z}}

\title{Expansive Automata Networks}

\author{Florian Bridoux\footnote{Universit\'e d’Aix-Marseille, CNRS, Centrale Marseille, LIS, Marseille, France. florian.bridoux@lis-lab.fr} \and
Maximilien Gadouleau\footnote{Department of Computer Science, Durham University, Durham, UK. m.r.gadouleau@durham.ac.uk} \and
Guillaume Theyssier\footnote{Universit\'e d’Aix-Marseille, CNRS, Centrale Marseille, I2M, Marseille, France. guillaume.theyssier@cnrs.fr}}

\begin{document}

\maketitle

\begin{abstract}
  An Automata Network is a map ${f:Q^n\rightarrow Q^n}$ where $Q$ is a finite alphabet. It can be viewed as a network of $n$ entities, each holding a state from $Q$, and evolving according to a deterministic synchronous update rule in such a way that each entity only depends on its neighbors in the network's graph, called interaction graph. A major trend in automata network theory is to understand how the interaction graph affects dynamical properties of $f$. In this work we introduce the following property called expansivity: the observation of the sequence of states at any given node is sufficient to determine the initial configuration of the whole network. Our main result is a characterization of interaction graphs that allow expansivity. Moreover, we show that this property is generic among linear automata networks over such graphs with large enough alphabet. We show however that the situation is more complex when the alphabet is fixed independently of the size of the interaction graph: no alphabet is sufficient to obtain expansivity on all admissible graphs, and only non-linear solutions exist in some cases. Finally, among other results, we consider a stronger version of expansivity where we ask to determine the initial configuration from any large enough observation of the system. We show that it can be achieved for any number of nodes and naturally gives rise to maximum distance separable codes.\footnote{This work was funded by the CNRS and Royal Society joint research project PRC1861.}
\end{abstract}

\section{Introduction}

Networks of interacting entities can be modelled as follows. The network consists of $n$ entities, where each entity $v$ has a local state represented by a $q$-ary variable $x_v \in \(q\) = \{0,1,\dots,q-1\}$, which evolves according to a deterministic function $f_v : \(q\)^n \to \(q\)$ of all the local states. More concisely, the configuration of the network is $x = (x_1,\dots, x_n) \in \(q\)^n$, which evolves according to a deterministic function $f = (f_1,\dots,f_n) : \(q\)^n \to \(q\)^n$. The function $f$, which encodes everything about the network, is referred to as an \textbf{Automata Network}, or simply network (the term Finite Dynamical Systems has also been applied for these networks). Automata networks have been used to model different networks, such as gene networks, neural networks, social networks, or network coding (see \cite{GR16} and references therein for the applications of Automata networks). They can also be considered as a distributed computational model with various specialized definitions like in \cite{WR79i,WR79ii}.
The architecture of an Automata network $f: \(q\)^n \to \(q\)^n$ can be represented via its \textbf{interaction graph} $\IG(f)$, which indicates which update functions depend on which variables. In other words, the interaction graph represents the underlying network of entities and their influences on one another. 
A major topic of interest is to determine how the interaction graph affects different properties of the network, such as the number of fixed points or images (see \cite{Gad18b} for a review of known results on the influence of the interaction graph). In particular, a stream of work aims to design networks with a prescribed interaction graph and with a specific dynamical property, such as a being bijective \cite{Gad18a}, or having many fixed points \cite{GR11}, or converging towards a fixed point \cite{GR16}.

In this paper, we introduce the concept of \textbf{expansive} networks. A network is expansive if the initial configuration of the network can be determined from the future temporal evolution of any local state. Formally, $f$ is expansive if it satisfies the equivalent conditions:

\begin{enumerate}
	\item \label{it:prop1}
	For any $v \in \{1, \dots, n\}$, there exists $T$ such that the function $(f_v(x), \dots, f_v^T(x)) : \(q\)^n \to \(q\)^T$ is injective.

	\item \label{it:prop2}
	For any $v \in \{1, \dots, n\}$ and any distinct $x, y \in \(q\)^n$, there exists $t\geq 1$ such that $f_v^t(x) \ne f_v^t(y)$.
\end{enumerate}

We are mostly interested in general results on automata networks per se, without any particular application in mind. Nonetheless, as mentioned above, automata networks are versatile and can be seen and used from different points of view. The concept of expansive automata networks introduced here is meaningful for several of them.

\emph{Dynamical systems.} The term 'expansive' is coined after the classical notion of dynamical system theory which corresponds to a strong form of topological unpredictability \cite{U50}. In the case of cellular automata \cite{kurkabook}, this topological notion has a concrete interpretation in terms of traces: the orbit of the whole system can be deduced from its temporal trace on a limited spatial region (there is  a similar notion in the field of symbolic dynamics \cite{BL97}). The definition above follows the same idea and in fact their is a precise correspondence between expansivity in cellular automata and expansivity in automata networks (see appendix~\ref{app:expca}). 

\emph{Distributed computation.} An expansive automata network can be seen as a protocol which solves the problem of giving the knowledge of the whole network's configuration to each of its entities. Moreover, if the constant $T$ in Property~\ref{it:prop1} above is optimal, \textit{i.e.} equal to $n$, then the automata network has another interesting algorithmic property: initial configurations ($n$ states) are mapped bijectively to sequences of states of length $n$ at each node. Said differently, it gives a protocol to transform a random initial configuration to a temporal random sequence of states at each node (with the uniform distribution in both cases). Such expansive networks with optimal constant $T$ exist as we show below.

\emph{Modeling tool and experimental sciences.} With this general point of view in mind, one is interested in making predictions on the future of a system from partial observations. In particular some components of the systems might be difficult or impossible to observe. Expansive automata network correspond to a favorable case where observing any single component for some time is sufficient. It is also interesting to relax the kind of observations allowed on the system, like in the stronger form of expansivity we consider in section~\ref{sec:strong}. 

\emph{Orthogonal arrays and maximum distance separable code.} In an expansive automata network, the orbit of any configuration contains a lot of redundancy because knowing $T$ consecutive states of any given node is sufficient to reconstruct the complete orbit. Pushing this idea further we obtain a stronger form of expansivty in section~\ref{sec:strong} which yields orthogonal arrays of index unity or equivalently maximum distance separable codes (see \cite{HSS99,MS77}). More precisely the set of orbits of a certain length is the code, and, in the linear case, it can be compactly represented by just giving the global map $f$ of the automata network.

\textbf{Our contributions.} First, in section~\ref{sec:exists}, we study the existence of expansive networks depending on the interaction graph. We characterize the graphs that admit an expansive network over some alphabet (Theorem~\ref{th:expansive}): those are the graphs that are strong (there is a path from any vertex to any other vertex) and coverable (the vertices can be covered by disjoint cycles). We show in particular that for such graphs, almost all linear networks over sufficiently large fields are expansive (Corollary~\ref{cor:randomlinear}). 

Second, we consider the existence of expansive networks over a given interaction graph and a given alphabet size. In section \ref{app:graphallalpha}, we exhibit two classes of graphs that admit expansive networks over all alphabets: the family of cycle with loops (Proposition~\ref{prop:cycle_with_loops}) and the family of cycles of cycles (Proposition~\ref{prop:cycle_of_cycles}). Conversely, in section~\ref{sec:nonexists}, we focus on the non-existence of expansive networks over small alphabets. We show that for any fixed alphabet size $q$, there exist strong and coverable graphs that do not admit any expansive network over $\(q\)$ (Theorem~\ref{th:no_expansive}). We also exhibit a graph which admits an expansive network over all alphabets, but does not admits any linear expansive network for an infinite number of alphabet values (Proposition~\ref{prop:onlynonlinear}). 

Then, in section~\ref{sec:exptime}, we focus on the minimum time $T(f)$, referred to as the expansion time, for which any difference between distinct configurations has been witnessed on all vertices. We notably prove that the expansion time can vary from $n$ to almost $q^n$, and that the minimum of $n$ is achieved by linear networks over fields (Theorem~\ref{th:expansion_time}). Moreover, in section~\ref{sec:expfreq}, we consider the average number of differences between two orbits, called expansion frequency. We show that it can be arbitrarily close to $1$ (Theorem~\ref{th:frequency}) while previous section gave a construction showing that it can be arbitrarily close to $0$. 

In section~\ref{sec:strong}, we consider a stronger notion of expansivity which asks to recover the initial configuration from any large enough observation of the system (not only the trace at a given node). We show that automata networks with that property yield maximum distance separable codes (Proposition~\ref{prop:supermds}) and exist on any complete interaction graph (Theorem~\ref{th:exists_super}), while they require an alphabet quadratic in the number of nodes (Corollary~\ref{cor:superbound}).

The concluding section~\ref{app:expca} discusses the relation of our notion of expansivity to that given for cellular automata.

\section{Definitions and preliminary results}

\textbf{Graphs.} A (directed) graph is a pair $D = (V,E)$, where $E \subseteq V^2$. For concepts about graphs, the reader is referred to the authoritative book \cite{BG09a}. Let us simply highlight some concepts and their notation in this paper. 
For any graph $D = (V,E)$ and any set of vertices $S \subseteq V$, we denote the out-neighbourhood of $S$ as $\NOut(S) = \{ u \in V : \exists s \in S \text{ s.t. } su \in E \}$; the in-neighbourhood is defined similarly and is denoted as $\NIn(S)$. An arc of the form $uu$ for some $u \in V$ is called a loop. A graph is loop-full if there is a loop on each vertex. For any two vertices $u,v \in V$, the distance from $u$ to $v$ in $D$ is the length of a shortest path from $u$ to $v$ in $D$; it is denoted as $d_D(u,v)$. The adjacency matrix of $D$, denoted $A_D$ is the $n \times n$ matrix $A_D$ with $A_D(i,j) = 1$ if $ij \in E$ and $A_D(i,j) = 0$ otherwise.

\noindent\textbf{Automata networks.} Let $n$ be a positive integer and $q$ be an integer no less than $2$. We denote $[n] = \{1, \dots, n\}$ and $\(q\) = \{0, \dots, q-1\}$. A state is any element $x = (x_1, \dots, x_n) \in \(q\)^n$. For any $S = \{s_1, \dots, s_k\} \subseteq [n]$, we denote $x_S = (x_{s_1}, \dots, x_{s_k})$; the order in which these indices occur will not matter usually. We further denote $x_{-S} = x_{[n] \setminus S}$ and $x_{-i} = x_{[n] \setminus \{i\}}$.
We denote the set of functions $f : \(q\)^n \to \(q\)^n$ as $\functions(n,q)$. A network is any element of $\functions(n,q)$. We can view $f$ as $f = (f_1, \dots, f_n)$, where each $f_v$ is a function $\(q\)^n \to \(q\)$. We can then use the same shorthand notation as for states, and define $f_S$ and $f_{-S}$, for instance. We also often use the notation $f_v^t=(f^t)_v$.
The interaction graph of $f \in \functions(n,q)$ has vertex set $V = [n]$ and has an arc from $u$ to $v$ if and only if $f_v$ depends essentially on $v$, i.e. there exists $a,b \in \(q\)^n$ such that $a_{-u} = b_{-u}$ and $f_v(a) \ne f_v(b)$. If the interaction graph of $f$ is $D$, we then say that $D$ admits $f$. We denote the set of networks in $\functions(n,q)$ with interaction graph $D$ as $\functions[D,q]$.

\noindent\textbf{Linear networks.} We shall focus on networks of a special kind; we give them in decreasing order of generality.
\begin{enumerate}
	\item \label{it:abelian}
	A network is \textbf{abelian} if $\(q\)$ is endowed with the structure of an abelian group $A$ and $f$ is an endomorphism of the group $A^n$. More concretely, we have
${f_v(x) = \sum_{j \in [n]} e_{v,j} (x_j)}$,
	where the $e_{v,j}$ are endomorphisms of $A$.
	
	\item \label{it:linear}
	A network is \textbf{linear} if $\(q\)$ is endowed with a ring structure $R$ and $f(x) = xM$, where $M \in R^{n \times n}$.
	
	\item \label{it:field}
	A network is \textbf{field linear} if it is linear over the finite field $\GF(q)$ of order $q$.
	
	\item \label{it:XOR}
	The \textbf{XOR network} on $D$ is $f \in \functions[D,2]$, defined by $f(x) = x A_D$, where $A_D$ is the adjacency matrix of $D$. This is the only abelian network with interaction graph $D$ for $q=2$.
\end{enumerate}

\noindent\textbf{Trace and expansive networks.} Fix $f \in \functions(n,q)$. Then for any $x$ and $v$, the \textbf{trace} of $x$ at $v$ is the infinite sequence ${\rho_v(x) = f_v(x), f_v^2(x), \dots}$.
We also denote $\rho_v^{(T)}(x) = (f_v(x), \dots, f_v^T(x))$ as the first $T$ elements in the trace.
A network is \textbf{expansive} if for any distinct $x, y \in \(q\)^n$ and any $v \in [n]$, there exists $t\geq 1$ such that $f_v^t(x) \ne f_v^t(y)$. Equivalently, $f$ is expansive if and only if for any $v$, there exists $T \ge 1$ such that the trace function $\rho_v^{(T)}$ is injective.
If $f$ is abelian, the function $(f_v, f^2_v, \dots, f^T_v) : \(q\)^n \to \(q\)^T$ is also abelian. Therefore, if $f$ is abelian then $f$ is expansive if and only if for all $x \in \(q\)^n \setminus\{ (0, \dots, 0) \}$ and all $v \in [n]$, there exists $t \ge 1$ such that $f_v^t(x) \ne f_v^t(0, \dots, 0)$. 
Let $f$ be a linear network, i.e. $f(x) = xM$. From $M$, construct the powers of $M$: $M^0 =I, M^1 = M, M^2, \dots$; denote the $u$-th column of $M^i$ as $M_u^i$. For any $t \ge 0$ and any $u \in [n]$, construct the matrix
\[
	N^{(t)}_u = \left( \begin{array}{c|c|c|c} M^t_u & M^{t+1}_u & \dots & M^{t + n - 1}_u \end{array} \right).
\]
The matrices $N^{(t)}_u$ then determine whether $f$ is expansive when $f$ is field linear.
\begin{lemma} \label{lem:expansive_linear}
The following are equivalent for a field linear network $f(x) = xM$.
\begin{enumerate}
	\item \label{it:1} $f$ is expansive.
	
	\item \label{it:2} $M$ is nonsigular and $N_u^{(0)}$ is nonsingular for all $u \in [n]$.
	
	\item \label{it:4} $N_u^{(t)}$ is nonsingular for all $u \in [n]$ and $t \ge 1$.
	
	\item \label{it:3} There exists $t \ge 1$ such that $N_u^{(t)}$ is nonsigular for all $u \in [n]$.
\end{enumerate}
\end{lemma}
\begin{proof}
We prove \ref{it:2} implies \ref{it:4}. Since $M_u^{t+k} = M^t M_u^k$, we have $N^{(t)}_u = M^t N_u^{(0)}$. Thus, if $M$ and $N_u^{(0)}$ are nonsingular, then so is $N_u^{(t)}$.
Clearly, \ref{it:4} implies \ref{it:3}. 
We prove \ref{it:3} implies \ref{it:1}. Suppose \ref{it:3} holds, and let $y = (f^t_u(x), \dots, f_u^{t+n-1}(x)) = x N_u^{(t)}$. Then $x = y (N_u^{(t)})^{-1}$ can be recovered from $y$ and $f$ is expansive.
We prove \ref{it:1} implies \ref{it:2}. Clearly, if $f$ is expansive, then it is bijective, thus $M$ is nonsingular. Suppose that $f$ is expansive, but $N_u^{(0)}$ is singular for some $u$. By expansivity, there exists $s > n$ such that the matrix
\[
	\tilde{N}^{(s)}_u = \left( \begin{array}{c|c|c|c} M^0_u & M^1_u & \dots & M^{s-1}_u \end{array} \right)
\]
is full rank. {However, because $N^{(0)}_u = \tilde{N}^{(n-1)}_u$ is singular}, there exists $j < n$ such that $M^j_u$ is in the column span of $\tilde{N}^{(j)}_u$. Say $M_u^j = \sum_{i=0}^{j-1} y_i M_u^i$, then we have for all $k \ge 0$
\[
	M_u^{j+k} = M^k M_u^j = M^k \sum_{i=0}^{j-1} y_i M_u^i = \sum_{i=0}^{j-1} y_i M_u^{i+k}.
\]
Thus, {$\rank(\tilde{N}^{(s)}_u) = \rank(\tilde{N}^{(j)}_u) \le \rank(N_u^{(0)}) < n$}, which is the desired contradiction.
\end{proof}

Since computing the determinant is no harder than multiplying matrices \cite[Theorem 6.6]{AHU74}, Property~\ref{it:2} (or Property~\ref{it:3} for $t=1$) yields an efficient algorithm to determine the expansivity of a field linear network. 
\begin{corollary}
Determining whether a field linear network is expansive can be done in $\mathcal{O}(n \cdot M(n))$, where $M(n)$ is the running time of an $n \times n$ matrix multiplication algorithm.
\end{corollary}

\section{Interaction graphs of expansive networks}
\label{sec:exists}

In this section, we are interested in determining for which interaction graphs $D$ there exists an expansive network in $\functions[D,q]$.

\noindent\textbf{Bijective networks.} Since an expansive network is bijective, we first derive a result about the existence of bijective networks. A cycle decomposition of a graph $D$ is a set of vertex-disjoint cycles that partition the vertex set of $D$. Say a graph $D$ is \textbf{coverable} if it has a cycle decomposition. A digraph is coverable if and only if $|\NOut(S)| \ge |S|$ for all $S \subseteq V$ \cite[Proposition 3.11.6]{BG09a}. These graphs can be characterized through existence of bijective networks.

\begin{theorem}[\cite{Gad18a}] \label{th:coverable}
  $D$ is coverable if and only if $\functions[D,q]$ contains a bijection for all $q \ge 3$.
\end{theorem}
We first give an analogue of the result on bijections in the linear case. Let $\matrices[D,q]$ denote the set of matrices over $\Z_q$ and with interaction graph equal to $D$:  ${M_{i,j} \ne 0 \iff ij \in E(D)}$.

\begin{theorem} \label{th:coverable_linear}
If $D$ is coverable, then $\matrices[D,q]$ contains a nonsingular matrix for any $q \ge 3$.
\end{theorem}

\begin{proof}
Let us first settle the case where $D$ is loop-full. We shall prove that there exists a matrix $M \in \matrices[D,q]$ with determinant equal to $1$. The result is clear for $n = 1$, so suppose it holds for $n-1$. Let $M' \in \matrices[D \setminus n, q]$ have determinant $1$. Then let $A \in \matrices[D,q]$ such that
\[
	A_{i,j} = \begin{cases}
	M'_{i,j}	&\text{if } i,j \ne n\\
	1			&\text{if } i=n \text{ and } j \ne n \text{ and } ij \in E, \text{or if } i \ne n \text{ and } j =n  \text{ and } ij \in E,\\
	1 			&\text{if } i=j=n\\
	0			&\text{otherwise}, 
	\end{cases}
\]
If $\det(A) \ne 2$, then consider $M \in \matrices[D,q]$ such that 
\[
	M_{i,j} = \begin{cases}
	2 - \det(A) &\text{if } i=j=n,\\ 
	A_{i,j}	&\text{otherwise}.
	\end{cases}
\]
Then {$\det(M) = \det(A) + (M_{n,n} - 1) \det(M') = 1$}. If $\det(A) = 2$, then let $M \in \matrices[D,q]$ such that 
\[
	M_{i,j} = \begin{cases}
	-2 	&\text{if } i=j=n,\\ 
	A_{i,j} &\text{if } j=n, i \ne n,\\
	-A_{i,j}	&\text{otherwise}.
\end{cases}
\]
We then have $\det(M) = \det(A) - \det(M') = 1$.

In the general case, let $D$ be coverable, then the mapping $v \mapsto \pi(v)$, where $\pi(v)$ is the successor of $v$ on a cycle in the cycle partition is a permutation. Denoting the permutation matrix of $\pi$ by $P$, define the graph $D'$ with adjacency matrix $A_{D'} = P^{-1} A_D$. Then $D'$ is loop-full, thus there exists $M' \in \matrices[D',q]$ with determinant $\text{sign}(\pi)$. Finally, the matrix $M := PM'$ belongs to $\matrices[D,q]$ and has determinant $1$.
\end{proof}

Recall that the term rank of a matrix is the maximum number of entries which are not in the same row or column. By the max-flow min-cut theorem, this is equal to the number of lines (rows and columns) necessary to cover all non-zero entries of the matrix. The term rank of the adjacency matrix of a graph is equal to the maximum number of pairwise independent arcs, where $uv, u'v'$ are independent if and only if $u \ne u'$ and $v \ne v'$; it is denoted as $\alpha_1(D)$ in \cite{Gad18a}.

\begin{corollary}[Edmonds's theorem \cite{Edm67}]
The maximum rank of a real matrix with interaction graph $D$ is equal to $\alpha_1(D)$. Moreover, the maximum is achieved by a matrix with entries in $\Z$.
\end{corollary}

\begin{corollary}[Theorem in \cite{Gad18a}]
The maximum rank of a network in $\functions[D,q]$ is equal to $q^{\alpha_1(D)}$ for all $q \ge 3$. Moreover, the maximum is achieved by a linear function over $\Z_q$.
\end{corollary}

\noindent\textbf{Expansive networks.} Some graphs admit expansive network for any alphabet (see appendix~\ref{app:graphallalpha}), some admit no expansive network whatever the alphabet.
We now characterize the graphs $D$ which admit an expansive network over some alphabet. We can actually be more precise, and consider variations of our main definition without affecting the characterization.
$f$ is said \textbf{weakly expansive} if for all $x \ne y$ and all $v$, there exists $t \ge 0$ such that $f_v^t(x) \ne f^t_v(y)$. Note that a weakling expansive $f$ has not to be bijective. $f$ is \textbf{quasi-expansive} if for all $x \ne y$ and all $v$, there exists $t \ge 0$ such that $f_{\NIn(v)}^t(x) \ne f^t_{\NIn(v)}(y)$. This last definition is the one corresponding to cellular automata as shown in appendix~\ref{app:expca}. It is not difficult to see that these definitions are not equivalent. However, the interaction graphs they characterize are the same as shown in the following theorem.

\begin{theorem} \label{th:expansive}
The following are equivalent for a graph $D$ on $n \ge 2$ vertices.
\begin{enumerate}
	\item $D$ is strong and coverable.

        \item $D$ admits an expansive network over some $q$.

	\item $D$ admits a quasi-expansive network over some $q$.

	\item $D$ admits a weakly expansive network over some $q$.

	\item $D$ admits a linear expansive network over any large enough finite field.
\end{enumerate}
\end{theorem}

Clearly, an expansive network is quasi-expansive and a weakly expansive. Therefore this theorem follows from the next 3 results.

\begin{lemma} \label{lem:necessary_expansive}
If $D$ admits a quasi-expansive network, then $D$ is strong and coverable (or $D = K_1$).
\end{lemma}
\begin{proof}
Let ${f\in\functions[D,q]}$ for some $q\geq 2$.
  
If $D$ is not strong, then let $u$ and $v$ such that there is no path from $u$ to $v$ in $D$. There is no path from $u$ to $\NIn(v)$ either. Then it is clear that for any $t \ge 0$, $f_W^t$ does not depend on $x_u$, where ${W=\NIn(v)}$. In particular, if $x,y \in \(q\)^n$ only differ in position $u$, we have $f_W^t(x) = f_W^t(y)$ for all $t \ge 0$. Therefore, $f$ cannot be quasi-expansive.

Suppose $D$ is not coverable, then by \cite[Proposition 3.11.6]{BG09a} there exists $S \subseteq V$ such that $|\NOut(S)| < |S|$. Choose any vertex ${v\not\in \NOut(S)}$ ($v$ may be in $S$ or not). By the pigeonhole principle there must exist two distinct configurations ${x,y\in \(q\)^n}$ such that $x$ and $y$ differ only on ${S}$ and ${f(x)_{\NOut(S)}=f(y)_{\NOut(S)}}$. But it also holds that ${f(x)_i=f(y)_i}$ for any ${i\not\in \NOut(S)}$ because ${x_{\NIn(i)}=y_{\NIn(i)}}$ since ${\NIn(i)\cap S=\emptyset}$. We deduce that ${f^t(x)=f^t(y)}$ for all ${t\geq 1}$ and ${x_{\NIn(v)}=y_{\NIn(v)}}$ which proves that $f$ is not quasi-expansive.
\end{proof}

\begin{lemma} \label{lem:necessary_non-bijective-expansive}
If $D$ admits a weakly expansive network, then $D$ is strong and coverable (or $D = K_1$).
\end{lemma}

\begin{proof}
The proof is similar to that of Lemma~\ref{lem:necessary_expansive}. Again, it is clear that $D$ must be strong. If $D$ is not coverable, then let $S \subseteq V$ such that $|\NOut(S)| < |S|$. If $S = V$, then any vertex $u$ outside of $\NOut(V)$ is a source, thus $f_u(x) = \text{cst} = c_u$. If we choose $x \ne y$  such that $x_u = y_u = c_u$, then we have $f^t_u(x) = f^t_u(y) = c_u$ for all $t \ge 0$. If $S \ne V$, then  there exist distinct configurations $x,y$ such that $x_S \ne y_S$ and ${x_{-S}=y_{-S}}$  and hence $f(x) = f(y)$. Thus, for any $v \notin S$ and any $t \ge 0$, $f^t_v(x) = f^t_v(y)$.
\end{proof}

\begin{theorem} \label{th:expansive_linear}
Any strong and coverable graph $D$ on $n$ vertices admits an expansive linear network over $\GF(q)$ for any prime power $q \ge \frac{1}{2} ( n^3 + n^2 + 4 )$.
\end{theorem}
\begin{proof}
We recall that a linear function $f(x) = x M$ is expansive if and only if for all $u \in [n]$, the matrix
\[
	N := N_u^{(1)} = \left( \begin{array}{c | c | c | c}
	M_u & M_u^2 & \dots & M_u^n
	\end{array} \right)
\]
is nonsingular. Our proof is nonconstructive: we shall see the nonzero coefficients of the matrix $M$ as variables, then the determinant of $N$ is a polynomial of these variables; if the field is large enough, then we can always evaluate that polynomial to something other than zero, provided it is not the null polynomial. 

Let $\bar{C}_1, \dots, \bar{C}_s$ be a decomposition of the vertex set of $D$ into cycles. We let $X(e) = \bar{\alpha}_k$ if $e$ is one of the arcs in $\bar{C}_k$; otherwise, we give a different variable $X(e) = \bar{\beta}_e$ for any other arc $e$ (and in particular for the chords of the cycles $\bar{C}_1, \dots, \bar{C}_s$). For any walk $W = e_1, \dots, e_L$ on $D$, we denote the monomial $X(W) = X(e_1)X(e_2) \cdots X(e_L)$. (The sum and the product of variables commute.)

We fix a vertex $u$, say it belongs to $\bar{C}_\sigma$. Let $T$ be a spanning ``tree of cycles'' rooted at $\bar{C}_\sigma$. More precisely, $T$ is a spanning subgraph of $D$ which contains all the cycles $\bar{C}_1, \dots, \bar{C}_s$ and for any $k \ne \sigma$, there is exactly one arc leaving $\bar{C}_k$. (Such a tree of cycles can be easily constructed by contracting every cycle to a vertex and then building a spanning in-tree rooted at the vertex corresponding to $\bar{C}_\sigma$.) It will be convenient to re-order the cycles according to the topological order in $T$. We then have $C_1 = \bar{C}_\sigma, C_2, \dots, C_s$. We similarly re-define the variables: $\alpha_k$ is the variable for all the arcs in $C_k$ ($1 \le k \le s$), while $\beta_k$ is the variable corresponding to the arc leaving $C_k$ in $T$ ($2 \le k \le s$). 

For $1 \le k \le s$, let $L_k$ be the length of $C_k$ and $\Lambda_k := L_1 + \dots + L_{k-1}$ ($\Lambda_1 = 0$). We also denote the shortest path from $C_k$ to $u$ in $T$ as $W_k$ and we denote its length as $\lambda_k$ and its monomial as $X_k = X(W_k)$; for $k = 1$, $W_1$ is the empty path thus $\lambda_1 = 0$ and $X_1 = 1$. It is easily seen that $\Lambda_k \ge \lambda_k$ for all $k$. We remark that for any distinct $v, v' \in C_k$, $d_T(v, u) \not\equiv d_T(v', u) \mod L_k$, where $d_T$ denotes the distance in $T$. We can then denote the vertices of $C_k$ according to their distance to $u$ as follows: let the vertices in $C_k$ be $v_k^j$ for $j=1, \dots, L_k$, where 
\[
	d_T( v_k^j, u ) \equiv \Lambda_k + j \mod L_k.
\]

For any row (vertex) $v$ and column (time) $t$, we have $N(v,t) = \sum_W X(W)$, where the sum is taken over all walks from $v$ to $u$ of length $t$. Let us consider $v = v_k^j$ and $t= \Lambda_k + j$. There is a canonical walk from $v$ to $u$ of time $t$: going round the cycle $C_k$ as many times as possible and then take the shortest path from $C_k$ to $u$, which yields the term $\alpha_k^{t - \lambda_k} X_k$. All the other walks either remain in $T$, but if so do not use $\alpha_k$ as many times, or leave $T$. This yields:
\[
	N( v_k^j, \Lambda_k + j ) = \alpha_k^{\Lambda_k + j - \lambda_k} X_k + \Gamma + \Delta,
\]
where all the terms in $\Gamma$ contain a variable outside of those of $T$, and the degree of $\alpha_k$ in $\Delta$ is at most $\Lambda_k + j - 1$. Therefore, the product
\[
	\prod_{\substack{ 1 \le k \le s\\ 1 \le j \le L_k }} N( v_k^j, \Lambda_k + j )
\]
contains the monomial
${Y := \prod_{k = 1}^s \alpha_k^{ d_k } X_k^{L_k}}$,
where
${d_k :=  L_k \left( \frac{1}{2}(L_k + 1) + \Lambda_k - \lambda_k \right)}$.

The term $Y$ contributes to the determinant of $N$, for the permutation $\rho$ of $[n]$, defined as $\rho( v_k^j) = \Lambda_k + j$. We now prove that $Y$ does not appear in any other product of entries that contribute to the determinant of $N$. More precisely, we prove by induction on $k$ from $s$ down to $1$ that if any permutation $\pi$ of $[n]$ produces a term only involving variables from $T$ where $\alpha_k$ has degree $d_k$, then $\pi(v) = \rho(v)$ for all $v \in C_k$. 

Let us prove the case $k = s$. Let $\pi(C_s) = \{t_1, \dots, t_{L_s}\}$ with $t_1 < \dots < t_{L_s}$. Clearly, we only need to consider walks in $T$. According to the topological order of cycles in $T$, there is no path from $C_i$ to $C_j$ if $i < j$. In particular, the rows of $N$ corresponding to a vertex outside of $C_s$ does not contain $\alpha_s$. Thus, the degree of $\alpha_s$ is at most $(t_1 - \lambda_s) + \dots + (t_{L_s} - \lambda_s)$. We then have 
\[
	t_1 + \dots + t_{L_s} - L_s \lambda_s \ge d_s = L_s \left( \frac{1}{2}(L_s + 1) + \Lambda_s \right) - L_s \lambda_s ,
\] 
which implies $t_j = \Lambda_s + j$ for $j = 1, \dots, L_s$. Moreover, the degree of $\alpha_s$ in $N( v_s^i, \Lambda_s + j )$ is equal to $\Lambda_s + j - \lambda_s$ if and only if $i = j$. This implies that $\pi(v_s^j) = \Lambda_s + j$ for all $j$. The inductive step is similar and hence omitted.

We have thus shown that $\det(N_u^{(1)})$ is a nonzero polynomial in the variables $\{ X(e) : e \in E \}$. Its degree is clearly $d := n(n+1)/2$. By the Schwartz-Zippel Lemma \cite[Theorem 7.1.4]{MP13}, there are at most $d(q-1)^{|E| - 1}$ choices for the values of $X(e)$ for which $\det(N_u^{(1)}) = 0$. Thus, there are at most $nd(q-1)^{|E| - 1}$ choices for the values of $X(e)$ for which $\det(N_u^{(1)}) = 0$ for some $u \in [n]$. Since $q - 1 > nd$, we have $(q-1)^{|E|} > nd(q-1)^{|E| - 1}$, and hence there exists a choice of values for all the variables $X(e)$ such that $\det(N_u^{(1)}) \ne 0$ for all $u$.
\end{proof}

We highlight two consequences of our result. Firstly, we comment on the alphabets for which a strong and coverable $D$ admits a linear expansive network. The \textbf{cartesian product} of two networks $f \in \functions(n,q)$ and $g \in \functions(n,r)$ is defined as follows. We view $\(qr\) \cong \(q\) \times \(r\) = \{ (a^1, a^2)  : a^1 \in \(q\), a^2 \in \(r\) \}$, then $f \times g = h \in \functions(n, qr)$ with $h(x^1,x^2) = ( f(x^1), g(x^2) )$. Some properties of the cartesian product are listed below; their proof is straightforward.
\begin{enumerate}
	\item If $f$ and $g$ are expansive, then so is $f \times g$.
	
	\item If $f$ and $g$ are linear, then so is $f \times g$.
	
	%\item $T(f \times g) = \max\{ T(f), T(g) \}$. Therefore, if $f$ and $g$ are strongly expansive, then so is $f \times g$.
	
	\item If $f$ and $g$ have interaction graph $D$, then so does $f \times g$.
\end{enumerate}
In particular, if $D$ admits a linear expansive network over alphabets of size $q$ and $r$, then it admits a linear expansive network over an alphabet of size $qr$.

\begin{corollary}
  \label{cor:density}
  For any $D$, the set of alphabet sizes $q$ for which there exists a linear expansive network in $\functions[D,q]$ has positive density.
\end{corollary}

\begin{proof}
Denote the prime numbers as $p_1 < p_2 < \dots$. Say $p_j$ is the largest prime no greater than $q := \frac{1}{2} ( n^3 + n^2 + 4 )$, then let $d_i = \lceil \log_{p_i} q \rceil$ for all $i \le j$. Then $D$ admits an expansive network on any multiple of $Q := \prod_{i=1}^j p_i^{d_i}$.
\end{proof}

\begin{conjecture}
For any strong and coverable $D$, there exists $q$ such that $D$ admits an expansive network over all alphabets of size greater than $q$.
\end{conjecture}

A corollary of Theorem~\ref{th:expansive_linear} is that choosing the coefficients $\alpha_k$ and $\beta_k$ at random will almost surely yield an expansive network when $q$ is large enough. Even more strikingly, we can define the following strategy to construct entire families of expansive networks. For a given $n$ and prime power $q$, the Random-Linear-Strategy first chooses a random matrix $M \in \GF(q)^{n \times n}$ whose entries are all nonzero. Then for a given graph $D$ on $[n]$, the strategy yields the linear network $f(x) = x (M \odot A_D)$ and $\odot$ denotes the Hadamard product of matrices.

\begin{corollary}

  \label{cor:randomlinear}
  The Random-Linear-Strategy produces an expansive network for all strong and coverable graphs on $n$ vertices with probability at least $1 - \Delta/(q-1)$, where $\Delta = 2^{n^2 - 1} n^2 (n+1)$.
\end{corollary}

\begin{proof}
Let $\alpha = \{ \alpha_{ij} : i,j \in [n] \}$ be an outcome of the Random-Linear-Strategy, where $\alpha_{ij}$ is a nonzero element of $\GF(q)$ for all $q$. For any strong and coverable graph $D$ on $n$ vertices, there are at most $n \cdot n(n+1)/2 \cdot (q-1)^{n^2-1}$ choices for $\alpha$ which do not yield an expansive network on $D$. Since there are at most $2^{n^2}$ choices for $D$, there are at most $\Delta (q-1)^{n^2-1}$ choices of $\alpha$ which do not produce an expansive network for all $D$. Thus, the probability of success is at least $1 - \Delta (q-1)^{n^2-1}/ (q-1)^{n^2}$.
\end{proof}

\section{Families of graphs with expansive networks over all alphabets}
\label{app:graphallalpha}

We exhibit two families of graphs which generalise the cycle, in the sense that the cycle belongs to either family and that every member of the family admits an expansive network over any alphabet (apart from one exception).

The first family is that of cycles with loops. Say $D = (V = [n], E)$ is a \textbf{cycle with loops} if there exists $S \subseteq [n]$ such that $E = \{ (i, i+1) : 1 \le i \le n \} \cup \{ (s,s) : s \in S \}$. Say a cycle with loops is proper if $S \ne V$.

We shall repeatedly use the following facts, whose proofs are obvious and hence omitted. Firstly, the following are equivalent:
\begin{enumerate}
	\item The XOR network on $D$ is bijective.
	
	\item The adjacency matrix $A_D$ is nonsingular over $\GF(2)$.
	
	\item $D$ has an odd number of cycle decompositions.
\end{enumerate}
Secondly, if $D$ has a unique cycle decomposition, then the adjacency matrix $A_D$ has determinant one over all rings $\Z_q$.

\begin{proposition} \label{prop:cycle_with_loops}
If $D$ is a cycle with loops, then $D$ admits an expansive linear network for any $q \ge 2$, unless $D$ is an improper cycle with loops and $q=2$.
\end{proposition}

\begin{proof}
We recall that a linear network $f(x) = x M$ is expansive if for all $u \in [n]$, the matrix $M$ and the matrix
\[
	N_u^{(0)} = \left( \begin{array}{c | c | c | c}
	I_u & M_u & \dots & M_u^{n-1}
	\end{array} \right)
\]
are nonsingular. If $D$ is a proper cycle with loops, then let $M = A_D$. Since $D$ has a unique cycle decomposition, $A_D$ is nonsigular. Also, $N_u^{(0)}$ is upper triangular with all ones on the diagonal, hence its determinant is equal to one and $N_u^{(0)}$ is nonsingular.

If $D$ is an improper cycle with loops, we see that it has exactly two cycle decompositions. As such, the XOR network is not bijective and $D$ does not admit a linear network for $q=2$. For $q = 3$, let $a \in \Z_q \setminus \{0,1\}$ be invertible, $b = 1 - a$ if $n$ is odd and $b = a + 1$ if $n$ is even, and
\[
	M = \begin{pmatrix}
	b		& a			& 0			& \cdots	& 0\\
	0 		& 1 		& 1 		& \cdots 	& 0\\
	\vdots 	& \vdots	& \ddots 	& \ddots	& \vdots\\
	0		& 0			& \cdots	& 1			& 1\\
	1 		& 0			& \cdots  	& 0			& 1
	\end{pmatrix},
\]
then $\det(M) = 1$. Once again, $N_u^{(0)}$ is upper triangular, and $\det(N_u^{(0)})$ is a power of $a$, which shows that $N_u^{(0)}$ is nonsingular.
\end{proof}

The second family is that of cycles of cycles. Say $D$ is a \textbf{cycle of cycles} if either it is a cycle or it is a union of $k \ge 2$ disjoint cycles $C_1, \dots, C_k$, linked as follows: for each cycle $C_i$ there are two vertices $u_i, v_i$ (which may be equal) such that $u_iv_{i+1} \in E$ for all $1 \le i \le k$ (computed cyclically). Say a cycle of cycles is proper if there exists $i$ such that $u_iv_i \notin E$.

\begin{proposition} \label{prop:cycle_of_cycles}
If $D$ is a cycle of cycles, then $D$ admits a linear expansive network for all $q \ge 2$ unless $D$ is an improper cycle of cycles and $q=2$.
\end{proposition}

\begin{proof}
Let $D$ be a proper cycle of cycles. Firstly, we verify that $D$ has a unique cycle decomposition. This is true when $D$ is a cycle. Otherwise, let $i$ such that $u_iv_i \notin E$, then the successor $w_i$ of $u_i$ belongs to only one cycle, namely $C_i$. Once $C_i$ is removed, it is then clear that $v_{i+1}$ only belongs to one cycle, namely $C_{i+1}$, and so on. Thus, the XOR network on $D$ is bijective. Conversely, if $D$ is an improper cycle of cycles, then it has exactly two cycle decompositions, and hence the XOR network is not bijective. For $q \ge 3$, there always exists a linear bijective network by Theorem~\ref{th:coverable_linear}.

Let $f$ be a linear bijective network on $D$. Suppose, for the sake of contradiction, that $f$ is not expansive. Let $x$ and $v$ such that for all $t \ge 0$, $f^t_v(x) = 0$. We consider two cases. Firstly, suppose that for all $1 \le i \le k$, there exists $t_i \ge 0$ such that $f^{t_i}_{u_i}(x) \ne 0$. Let $v$ be any vertex, say it belongs to $C_i$; denote the vertices of $C_i$ as $u_i, u_i + 1, \dots u_i + l_i - 1$ and in particular $v = u_i + b$. We then have
\[
	f^{t_i + a}_{u_i + a}(x) \ne 0
\]
for all $0 \le a \le l_i - 1$ and in particular $f^{t_i + b}_v(x) \ne 0$, which is the desired contradiction.

Secondly, suppose that there exists $1 \le j \le k$ such that $f^t_{u_j}(x) = 0$ for all $t \ge 0$. Then for all $t \ge l_j$, $f^t_{C_j}(x) = 0$. (Justify.) For any $t \ge l_j$, we have
\[
	0 = f^{t+1}_{v_j}(x) = f^t_{u_{j-1}}(x),
\]
thus by a similar reasoning, $f^t_{C_{j-1}}(x) = 0$ for all $t \ge l_j + l_{j-1}$. By obvious induction, we obtain that $f^n(x) = 0$, which contradicts the fact that $f$ is bijective.
\end{proof}

\section{Nonexistence of expansive networks}
\label{sec:nonexists}

We only consider strong and coverable graphs from now on. For any graph $D$, we denote the set of expansive (abelian expansive, respectively) networks in $\functions[D,q]$ as $E[D,q]$ ($EA[D,q]$, respectively).
Our nonexistence results are based on the following family of graphs. Consider for any $n\geq 2$ the graph ${G_n=(V_n = \{0,1,\ldots,n\},E_n)}$ where ${E_n = \{ (0,i), (i,0), (i,i) : 0 \le i \le n \}}$.

\begin{theorem} \label{th:no_expansive}
For all $q$, there exists a strong coverable graph $G$ such that $E[G,q] = \emptyset$.
\end{theorem}
\begin{proof}
  We shall prove that for any $q$ there is $n$ large enough such that ${E[G_n,q]=\emptyset}$. We first show that any ${f\in\functions[G_n,q]}$ has a lot of initial configurations reaching cycles of constant size in constant time (\textit{i.e.} independent of $n$). To make a precise statement, denote for any ${\phi\in \(q\)^{\(q\)^ 2}}$ the set of automata in $f$ whose update map is precisely $\phi$: ${V_\phi = \{i : 0<i\leq n\text{ and }f_i = \phi\}}$. Choose any $\phi$, any ${V\subseteq V_\phi}$ and define the configuration ${c_{\phi,V}\in \(q\)^n}$ by: 
  \[c_{\phi,V}(j) =
    \begin{cases}
      0&\text{ if } j\not\in V\\
      1&\text{ else.}
    \end{cases}
  \]
  We claim that the orbit under $f$ of any such ${c_{\phi,V}}$ has length at most ${p=q^{q^{q^2}+2}}$. Indeed, by induction on $t$, it holds that ${f^t(c_{\phi,V})_i = f^t(c_{\phi,V})_j}$ as soon as ${\{i,j\}\subseteq V_{\phi'}}$ for ${\phi'\neq\phi}$, or ${\{i,j\}\subseteq V}$, or ${\{i,j\}\subseteq V_{\phi}\setminus V}$ (it is true at $t=0$ and preserved because two automata ${\{i,j\}\subseteq V_{\phi'}}$ apply the same update map $\phi'$). It follows that there are at most ${q^{q^{q^2}+2}}$ different configurations in the orbit of ${c_{\phi,V}}$, which proves the claim.

  Now fix $q$, let ${p= q^{q^{q^2}+2}}$ and ${l>\lceil\log_2(q^{2p})\rceil}$ and choose ${n = lq^{q^2}}$. By choice of $n$ there must be ${\phi\in\(q\)^{\(q\)^ 2}}$ such that ${|V_\phi|\geq l}$. Thus, there are ${2^l}$ choices of ${V\subseteq V_\phi}$ yielding ${2^l}$ distinct configurations of the form ${c_{\phi,V}}$. For any configuration $c$, define $\rho(c) := \rho^{(2p)}_0(c)$ the trace of length $2p$ at node $0$. By choice of $l$ there must be ${V,V'\subseteq V_\phi}$ with ${V\neq V'}$ such that ${h(c_{\phi,V})=h(c_{\phi,V'})}$ (because $h$ can take only ${q^{2p}}$ different values). Anticipating the notation from section~\ref{sec:exptime}, we thus have two distinct configurations ${x =c_{\phi,V}}$ and ${y = c_{\phi,V'}}$ such that $\tau_0(x,y) \ge l_x + l_y$, which contradicts the fact that $\tau_v(x,y) < l_x + l_y$, as shown in the proof of Theorem~\ref{th:expansion_time}.
\end{proof}

We now prove that the bound on the smallest $n$ such that there exists $G$ on $n$ vertices with no expansive networks over $\(q\)$ can be significantly lowered if we only consider linear networks. We give a proof that actually holds for abelian networks with a quasi-polynomial bound.

\begin{theorem} \label{th:no_linear_expansive}
For any $q \ge 2$ and any ${n>q^{2\log(q)}}$ it holds $EA[G_n,q] = \emptyset$.
\end{theorem}

\begin{proof}
Let $N_q$ denote the maximum number of endomorphisms of an abelian group of order $q$. By the decomposition theorem of Abelian groups intro products of cyclic groups, one sees that an endomorphism is determined by its value on at most $\log(q)$ elements (elements equal to the generator on one component of the product and 0 on the others), thus ${N_q\leq q^{\log(q)}}$. Let $n > q^{2\log(q)} \geq N_q^2$. Let $A$ be an abelian group of order $q$ and $f \in \functions[G_n, q]$ be an endomorphism of $A^n$. Then there exist $i$ and $j$ such that $f_i(x) = g(x_i) + h(x_0)$, $f_j(x) = g(x_j) + h'(x_0)$ and $f_0(x) = e(x_i) + e(x_j) + h''(x_{-\{i,j\}})$ for some endomorphisms $g$ and $e$ of $A$. Consider a nonzero configuration $x$ such that $x_i + x_j = 0$ and $x_u = 0$ for any other vertex $u$, we then have
\begin{align*}
	f_i(x) + f_j(x) &= g(x_i) - g(x_j) = 0,\\
	f_0(x)&= e(x_i) - e(x_j) = 0.
\end{align*}
By induction, we have $f^t_0(x) = 0$, thus $f$ is not expansive.
\end{proof}

The proof can be easily adapted for linear networks, thus yielding a polynomial bound on the smallest $n$ for which $G_n$ admits no linear network over an alphabet of size $q$.

\begin{corollary}
The graph $G_n$ admits no linear expansive network for $q$ whenever $n > (q-1)^2$.
\end{corollary}

We conjecture that in fact, there is a sharp distinction between admitting an expansive network and admitting an abelian expansive network.

\begin{conjecture} \label{con:EA}
For all $q$, there exists $D$ such that $E[D,q] \ne \emptyset$ but $EA[D,q] = \emptyset$.
\end{conjecture}

We make some progress towards Conjecture~\ref{con:EA} by showing that it holds for all $q \equiv 2 \mod 4$. Let $G$ be the graph on four vertices displayed below:

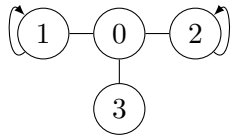
\begin{center}
\begin{tikzpicture}
	\node[draw, shape=circle](1) at (0,1) {$1$};
	\node[draw, shape=circle](2) at (2,1) {$2$};
	\node[draw, shape=circle](0) at (1,1) {$0$};
	\node[draw, shape=circle](3) at (1,0) {$3$};
	
	\draw	(0) -- (1); 
	\draw	(0) -- (2);
	\draw	(0) -- (3);
	\draw[-latex]	(1) .. controls	(-0.5, 0.5)	and (-0.5, 1.5) .. (1);
	\draw[-latex]	(2) .. controls	(2.5, 0.5)	and (2.5, 1.5) 	.. (2);
\end{tikzpicture}
\end{center}

\begin{proposition} \label{prop:onlynonlinear}
We have $E[G,q] \ne \emptyset$ for all $q \ge 2$ and $EA[G,q] \ne \emptyset$ if and only if $q \not\equiv 2 \mod 4$.
\end{proposition}

\begin{proof}
Firstly, we verify that $G$ admits no abelian expansive network for $q=2$. For $q=2$, the only abelian network is the XOR network $f(x) = x A_G$. The configuration $x = (0,1,1,0)$ is a fixed point of the XOR network, thus the latter is not expansive. More generally, for any $q = 2k$ for $k$ odd, any abelian network $h \in \functions(n,2k)$ decomposes as $h = f \times g$, where $f \in \functions(n,2)$ and $g \in \functions(n,k)$ are both abelian. We thus obtain that $G$ admits no abelian expansive network for any $q \equiv 2 \mod 4$.

Secondly, we show that there exists an abelian expansive over $G$ for all $q \not\equiv 2 \mod 4$. We only need to prove the case for non-binary finite fields, the general case following by cartesian product. Let $q \ne 2$ be a prime power and let $\alpha \ne \{0,1\}$ be an element of $\GF(q)$. Let $f(x) = x M$, where
\[
	M = \begin{pmatrix}
	0 & 1 & 1 & 1 \\
	1 & 1 & 0 & 0 \\
	1 & 0 & \alpha & 0\\
	1 & 0 & 0 & 0
	\end{pmatrix}.
\]
Clearly, $\det(M) = -\alpha$ and hence $f$ is bijective. After some straightforward calculations, we obtain
\begin{alignat*}{2}
	N_0 &= \begin{pmatrix}
	1 & 0 & 3 & \alpha + 1\\
	0 & 1 & 1 & 4\\
	0 & 1 & \alpha & \alpha^2 + 3\\
	0 & 1 & 0 & 3
	\end{pmatrix},
	&\qquad
	\det(N_0) &= \alpha^2 - \alpha;\\
	N_1 &= \begin{pmatrix}
	0 & 1 & 1 & 4\\
	1 & 1 & 2 & 3\\
	0 & 0 & 1 & \alpha + 1\\
	0 & 0 & 1 & 1
	\end{pmatrix},
	&\qquad
	\det(N_1) &= \alpha;\\
	N_2 &= \begin{pmatrix}
	0 & 1 & \alpha & \alpha^2 + 3\\
	0 & 0 & 1 & \alpha + 1\\
	1 & \alpha & \alpha^2 + 1 & \alpha^3 + 2 \alpha\\
	0 & 0 & 1 & \alpha
	\end{pmatrix},
	&\qquad
	\det(N_2) &= -1;\\
	N_3 &= \begin{pmatrix}
	0 & 1 & 0 & 3\\
	0 & 0 & 1 & 1\\
	0 & 0 & 1 & \alpha\\
	1 & 0 & 1 & 0
	\end{pmatrix},
	&\qquad
	\det(N_3) &= 1 - \alpha.
\end{alignat*}
All determinants are nonzero, thus $f$ is expansive.

Thirdly, for $q=2$, it is straightforward to check that the following network is indeed expansive.
\begin{align*}
	f_0(x) &= x_1x_2 + x_3 + 1\\
	f_1(x) &= x_0 + x_1\\
	f_2(x) &= x_0 + x_2 + 1\\
	f_3(x) &= x_0.
\end{align*}
Again, combining our previous results and using the cartesian product, we conclude that $G$ admits an expansive network for all $q \ge 2$.
\end{proof}

\section{Expansion time}
\label{sec:exptime}

Consider some expansive network $f$. For any node $v$ and any configuration $x$, it is clear that the trace ${\rho_v(x)}$ is periodic. In particular, let $O_x = \{ f(x), \dots, f^{l_x}(x) = x \}$ be the orbit of $x$, then the period of the trace of $x$ is equal to the size of its orbit $l_x$ (if it where shorter of length $l$, then $x$ and $f^l(x)$ would be two distinct configurations of same trace). 

For any expansive network $f \in \functions(n,q)$, any different $x, y \in \(q\)^n$ and any $v \in [n]$, let
\[
	\tau_v(x,y) = \min \left\{ t \ge 1 : f^t_v(x) \ne f^t_v(y)  \right\}.
\]
The \textbf{expansion time} of $f$ is then
\[
	T(f) := \max_{x \ne y \in \(q\)^n, v \in [n]} \tau_v(x,y).
\]
This is the shortest time for which the temporal evolution of $x_v$ determines $x$ completely, for any $x$ and any $v$.
For any $v$, it is clear that if the function $\rho_v^{(T)}$ is injective, then $T \ge n$, thus $T(f) \ge n$. Say $f$ is \textbf{strongly expansive} if it is expansive and $T(f) = n$. Lemma~\ref{lem:expansive_linear} then shows that any expansive field linear network is strongly expansive. Strongly expansive networks can be viewed as follows. For any $x \in \(q\)^n$, consider the matrix
\[
	M_x = \begin{pmatrix}
	f_1(x) & f_2(x) & \dots & f_n(x)\\
	f^2_1(x) & f^2_2(x) & \dots & f^2_n(x)\\
	\vdots & \vdots & \dots & \vdots\\
	f^n_1(x) & f^n_2(x) & \dots & f^n_n(x)
	\end{pmatrix}.
\]
Then $f$ is bijective if and only if we can recover $x$ from any \textit{row} of $M_x$, while $f$ is strongly expansive if and only if we can recover $x$ from any \textit{column} of $M_x$.

The expansion time is the maximum value of $T$ such that one can recover any $x$ from the first $T$ time steps of its trace at $v$. For a given $x$ and a given $v$, that particular time may be smaller than $n$ as shown in the following example.
 \begin{example}
 Let $f \in \functions(3,2)$ be defined as\\
 ~\\
 \begin{tabular}{c|c}
 	$x$ & $f(x)$\\
 	\hline
 	000 & 001 \\
 	001 & 110 \\
 	010 & 101 \\
 	011 & 111 \\
 	100 & 011 \\
 	101 & 010 \\
 	110 & 000 \\
 	111 & 100 
 \end{tabular}
 ~\\
 ~\\
 It can be checked that $f$ is indeed expansive, with expansion time $T(f) = 4$ (and hence $f$ is not strongly expansive). Then the traces at vertex $v = 1$ are as follows. We highlight the part of the trace that allows to recover the initial state.\\
 ~\\
 \begin{tabular}{c|c}
 	$x$ & $(f_1(x), f^2_1(x), f^3_1(x), f^4_1(x))$\\
 	\hline
 	000 & \textbf{0100} \\
 	001 & \textbf{100}1 \\
 	010 & \textbf{1010} \\
 	011 & \textbf{11}01 \\
 	100 & \textbf{011}0 \\
 	101 & \textbf{0101} \\
 	110 & \textbf{00}10 \\
 	111 & \textbf{1011}
 \end{tabular}
 ~\\
In particular, the states $011$ and $110$ could be recovered after only two time steps.
\end{example}

However, the expansion time is ``universal'' for strongly expansive networks: for any $v$ and any $x$, one must wait $n$ time steps before being able to recover $x$.

\begin{proposition} \label{prop:universaltime}
If $f$ is strongly expansive, then for any $v \in [n]$ and $x \in \(q\)^n$, there exists $y \ne x$ such that $\tau_v(x,y) = n$. 
\end{proposition}

\begin{proof}
If $f$ is strongly expansive, then for any $v$, the function $\rho_v^{(n)}$ is surjective. Suppose, for the sake of contradiction, that there exists $x$ such that for any $y \ne x$, $\rho_v^{(n-1)}(x) \ne \rho_v^{(n-1)}(y)$. Let $a \ne f^n_v(x)$, then there is no $y$ such that $\rho_v^{(n)}(y) = (\rho_v^{(n-1)}, a)$, thus contradicting surjectivity.
\end{proof}

In order to highlight the specificity of strongly expansive networks, we now show that the maximum possible expansion time is almost $q^n$.

\begin{theorem} \label{th:expansion_time}
For all $n$ and $q$, the maximum $T(f)$ over all expansive $f \in \functions(n,q)$ is between $q^n - q - 1$ and $q^n - 2$.
\end{theorem}
\begin{proof}
\textit{Upper bound.} Let $x \ne y$ with $l_x \le l_y$; denote $\tau = \tau_v(x,y)$.

Case 1: $l_x \,|\, l_y$. We first prove that $\tau \le l_y - 1$. Suppose that this is not the case, i.e. $f_v^t(x) = f_v^t(y)$ for all $1 \le t \le l_y$. Then $\rho_v(x) = \rho_v(y)$, which contradicts the expansivity of $f$. Thus, $\tau \le l_y - 1 \le q^n -1$. Suppose that $\tau = l_y-1 = q^n-1$, then $f$ is a cyclic permutation of $\(q\)^n$ and the trace $\rho_v(x)$ is a cyclic shift of $\rho_v(y)$, and their only difference is in position $q^n$, i.e $x_v = a \ne y_v = b$. Let $N = |\{  t : 1 \le t \le q^n - 1, f^t_v(x) = a \}|$ denote the number of times the trace of $x$ is equal to $a$ until time $q^n - 1$. We then have $N = q^{n-1} - 1$. However, $N$ also counts the number of times the trace of $y$ is equal to $a$ until time $q^n - 1$; we obtain $N = q^{n-1}$, which is the desired contradiction.

Case 2: $l_x \not | \, l_y$. We prove that $\tau \le l_x + l_y - \gcd(l_x, l_y) - 1$. Suppose, for the sake of contradiction, that $\tau \ge l_x + l_y - \gcd(l_x, l_y)$. We shall reason in terms of blocks of length $\gcd(l_x, l_y)$. Say the first period of the trace of $x$ is $X = u_1, \dots, u_{|X|}$ and that of $y$ is $Y = u_1, \dots, u_{|Y|}$ (this is coherent since $X$ is a prefix of $Y$). We then have $|X| = l_x / \gcd(l_x, l_y)$ and $|Y| = l_y / \gcd(l_x, l_y)$; these two are coprime. Let $|Y| = \alpha |X| + a$ for $0 \le a < |X|$ and $|X| = \beta a + b$ for $0 \le b < a$, then $a$ and $b$ are coprime. 

\begin{claim}
Let $u := u_1, \dots, u_a$ and $u' := u_1, \dots, u_b$. Then $X = u^\beta,u'$ and $Y = X^\alpha, u$.
\end{claim}

\begin{proof}
Clearly, we have $Y = X^\alpha, v$ for $v = u_{\alpha |X| + 1}, \dots, u_Y$. At times $\alpha |X| + 1$ to $\alpha |X| + a$, the trace of $x$ describes $u$, thus $v = u$. This proves the second claim. Similarly, at times $|Y| + 1 = \alpha |X| + a + 1$ to $|Y| + a = \alpha |X| + 2a$, the trace of $y$ describes $u$, thus $X$ begins with $u,u$. By easy induction, we prove that $u$ is repeated throughout $X$ and we obtain $X = u^\beta, u'$.
\end{proof}

We now focus on times from $t := |X| + |Y| - a - b + 1$ to $t + a + b - 1 = |X| + |Y|$. The trace of $x$ describes $u_1, \dots, u_b, u_1, \dots, u_{a-1}$, while the trace of $y$ describes $u_1, \dots, u_a, u_1, \dots, u_{b-1}$. For the times from $t + b + 1$ to $t + a$, we obtain $u_i = u_{i+b}$ for all $1 \le i \le a-b$; for the times from $t+a+1$ to $t + a + b - 1$, we obtain $u_j = u_{j - a + b}$ for all  $a - b + 1 \le j \le a - 1$. Since $b$ is coprime to $a$, it is easily checked that we obtain $u_1 = \dots = u_a$. Thus, $X = u_1, \dots, u_1$, which contradicts its period.

\medskip

\textit{Lower bound.} We now construct an expansive network in $\functions(n,q)$ with expansion time $T(f) \ge q^n - q - 1$. Intuitively, this network is the successor function of a particular enumeration of $\(q\)^n$, which can be viewed as a ``twisted lexicographic order.'' More formally, for any integer $0 \le a \le q^n - 1$, say $a = a_0 + a_1 q + \dots + a_n q^{n-1}$, let $x^a = (x^a_1, \dots, x^a_n) \in \(q\)^n$ be defined as $x^a_n = a_n$ and for $1 \le i \le n-1$,
\[
	x^a_i = \begin{cases}
		q-2 &\text{if } a_{i+1} = \dots = a_n = q-1 \text{ and } a_i = q-1\\
		q-1 &\text{if } a_{i+1} = \dots = a_n = q-1 \text{ and } a_i = q-2\\
		a_i &\text{otherwise}.
	\end{cases}
\] 
Then let $f(x^a) = x^{a + 1 \mod q^n}$. Clearly, $f$ is bijective.

We now prove that $f$ is expansive. We only need to show that for any $1 \le e \le \lfloor q^n/2 \rfloor$ and any $v \in [n]$, there exists $0 \le t \le q^n-1$ such that $x_v^t \ne x_v^{e+t}$. Let $k$ be the largest number such that $q^{k-1}$ divides $e$. For $1 \le v \le k$, we have $x_v^{q^n - 1 - e} = q-1$ and $x_v^{q^n - 1} = q-2$, while for $k+1 \le v \le n$, we have $x_v^{q^{v-1} - 1} = 0$ and $x_v^{q^{v-1} - 1 + e} \ne 0$. Finally, it is easily verified that $\tau_1(x^{q^n-1}, x^{q-1}) = q^n - q - 1$.
\end{proof}

\section{Expansion frequency}
\label{sec:expfreq}

In the previous section, we have shown that we may have to wait until $n$ time steps in order to differentiate a particular pair $x,y$ of distinct states. However, that difference may occur frequently after its first occurrence. In this section, we are then interested at how often we see a difference between the orbits of $x$ and $y$ at some given node $v$.

For all distinct $x,y \in \(q\)^n$ and all $v \in [n]$, let
\[
	\phi_v(x,y) := \frac{ \dH( \rho_v(x)^{(l_x l_y)}, \rho(y)^{(l_x l_y)} ) }{ l_x l_y },
\]
where $\dH$ denotes the Hamming distance. We then define the \textbf{expansion frequency} of $f$ as
\[
	\Phi(f) = \min_{x \ne y \in \(q\)^n, v \in [n]} \phi_v(x, y).
\]
It is clear that ${\frac{1}{T(f)} \le \Phi(f) \le 1}$.
However, $\Phi(f)$ itself can be as close to $1$ as possible.

\begin{theorem} \label{th:frequency}
For any expansive network $f$, 
${\Phi(f) \le \frac{ q^n - q } { q^n - 1 }}$.
Equality is achieved for all $n$ and all prime powers $q$ by some field linear network.
\end{theorem}
\begin{proof}
\textit{Upper bound.} Let $N$ be the product of all the orbit lengths under $f$, and consider the code $C = \{ \rho_v^{(N)}(x) : x \in \(q\)^n \}$ of length $N$ over $\(q\)$. Since $T(f) \le N$, all traces are distinct and hence $|C| = q^n$.

\begin{claim}
The minimum distance of $C$ is bounded by:
${\dmin(C) \le \delta := \frac{(q-1)N q^{n-1} }{ q^n - 1 }}$
\end{claim}

\begin{proof}
Berlekamp's generalisation of the Plotkin bound in \cite{Ber68} shows that for any code $C$ of length $N$ over $\(q\)$ and minimum distance $d$, we have ${|C| \le \frac{dq}{dq - N(q-1)},}$
provided the denominator is positive. Since $\delta q > N (q-1)$, we can use Berlekamp's generalisation of the Plotkin bound. In particular, if $\dmin(C) = d > \delta$, then 
${|C| \le \frac{ dq }{ dq - N (q-1) } < \frac{ \delta q }{ \delta q - N (q-1) } = q^n}$,
which is a contradiction. Thus, $d \le \delta$. 
\end{proof}

By definition, there exists a pair $x,y \in \(q\)^n$ such that $\dH( \rho_v^{(N)}(x), \rho_v^{(N)}(y) ) = \dmin(C)$. We obtain
${\phi_v(x,y) \le \frac{\delta}{ N } = \frac{(q-1) q^{n-1} }{ q^n - 1 }}$.

\textit{Achievability.} Consider the $q$-ary image of the mapping $\xi \mapsto \alpha \xi$ in $\GF(q^n)$, where $\alpha$ is a primitive element of the field. This is a linear function in $\functions(n,q)$, with $0$ as its unique fixed point. For any nonzero $x \in \GF(q^n)$, the orbit of $x$ contains all $q^n - 1$ nonzero elements of $\GF(q)^n$. Therefore, for any $x \ne 0$ and $v$, $\phi_v(x,0) = \frac{ (q-1) q^{n-1} }{ q^n - 1 }$.
\end{proof}

On the other extreme, the construction in the proof of Theorem~\ref{th:expansion_time} yields a network with expansion frequency of $2/q^n$.

\section{Stronger form of expansivity}
\label{sec:strong}

 The notion of expansivity considered so far asks to determine the initial configuration from the trace at any given node. Here, we strengthen the notion by asking to determine the initial configuration from any large enough 'observation' of the network during the first $n$ time steps.
 Let ${f\in\functions(n,q)}$. Consider any sequence $\omega$ of $n$ pairs (vertex, time step): ${\omega=\bigl(v_1,t_1),\ldots,(v_n,t_n)}$ where ${v_i\in[n]}$ and ${t_i\in[n]}$ for all ${1\leq i\leq n}$. The associated \emph{observation} is the map ${\tau_\omega:\(q\)^n\rightarrow \(q\)^n}$ given by ${\tau_\omega(x)=\bigl(f^{t_1}(x)_{v_1},f^{t_2}(x)_{v_2},\ldots,f^{t_n}(x)_{v_n}\bigr)}$. We say $f$ is \textbf{super-expansive} if for any $\omega$, the map $\tau_\omega$ is injective.
 Looking again at matrix $M_x$ defined previously, $f$ is super-expansive if $x$ can be determined from any set of $n$ entries in this matrix.

\begin{proposition} \label{prop:supercomplete}
  Let $D$ be a graph with $n$ nodes. If ${f\in\functions[D,q]}$ is super-expansive, then $D$ is the complete graph (the graph with $n^2$ arcs).
\end{proposition}

\begin{proof}
  Suppose that $D$ does not contain the arc ${ij}$ and consider any ${f\in\functions[D,q]}$. Let 
  \[
  	\omega = \bigl((1,1),\ldots,(i-1,1),(j,2),(i+1,1),\ldots,(n,1)\bigr),
  \]
  then the interaction graph of $\tau_\omega$ has a source (namely $i$) and hence is not coverable. Thus, by \cite{Gad18a}, $\tau_\omega$ is not bijective.
\end{proof}

Using a similar technique as in the proof of Theorem~\ref{th:expansive_linear} we can show the existence of super-expansive networks.

\begin{theorem} \label{th:exists_super}
  For any $n$ and any prime power $q> n^2{n^2 \choose n}$ there exists a super-expansive linear network with $n$ nodes over $\GF(q)$.
\end{theorem}

\begin{proof}
  The proof technique is similar to that of Theorem~\ref{th:expansive_linear}. First for any linear function ${f(x) = x M}$ and any ${\omega=\bigl(v_1,t_1),\ldots,(v_n,t_n)}$, the observation $\tau_\omega$ is injective if and only if the matrix
\[
	N_\omega := \left( \begin{array}{c | c | c | c}
	M_{v_1}^{t_1} & M_{v_2}^{t_2} & \dots & M_{v_n}^{t_n}
	\end{array} \right)
\]
is nonsingular (straightforward adaptation of Lemma~\ref{lem:expansive_linear}). Since injectivity of $\tau_\omega$ is preserved by permutation of $\omega$, we suppose that ${t_1\leq t_2\leq\cdots\leq t_n}$. Like for Theorem~\ref{th:expansive_linear} our proof is nonconstructive: we shall see the nonzero coefficients of the matrix $M$ as variables ${(X_{ij})_{i,j\in[n]}}$, then the determinant of $N_\omega$ is a polynomial of these variables; if the field is large enough, then we can always evaluate that polynomial to something other than zero, provided it is not the null polynomial. Using the correspondence between walks on the complete graph and monomials, the determinant of $N_\omega$ can be expressed as: 
\[\det(N_\omega)=\sum_{\sigma\in S_n}\epsilon(\sigma)P_\sigma\]
where each monomial appearing in $P_\sigma$ is of the form ${\prod_{i=1}^n\prod_{k=1}^{t_i}X_{w_i(k)}}$ where ${w_i(1)\cdots w_i(t_i)}$ is a walk of length $t_i$ from node $\sigma(i)$ to node $v_i$. We shall now choose a specific permutation $\sigma$ and a specific monomial appearing in $P_\sigma$, and show that it does not appear in any other $P_{\sigma'}$ for ${\sigma\neq\sigma'}$.
 Let ${A=\{v_1,\ldots,v_n\}}$ and choose ${v'_1,\ldots,v'_n}$ distinct nodes verifying: 
\[v'_i =
  \begin{cases}
    v_i&\text{ if $i=\min\{k:v_i=v_k\}$}\\
    \not\in A&\text{ else.}
  \end{cases}
\]
Our permutation is ${\sigma=i\mapsto v'_i}$, and we consider the monomial ${M=\prod_iX_{v'_iv'_i}^{t_i-1}X_{v'_iv_i}}$ which clearly appears in $P_\sigma$. Consider any permutation $\sigma'$ and suppose that $M$ appears in $P_{\sigma'}$, \textit{i.e.} 
\[
	M=\prod_{i=1}^n\prod_{k=1}^{t_i}X_{w_i(k)}
\]
where ${w_i(1)\cdots w_i(t_i)}$ is a walk of length $t_i$ from node ${\sigma'(i)}$ to node $v_i$. For each ${v\in A}$, let $I_v=\{i : v_i=v\}$. If ${I_v=\{i\}}$ is a singleton then we must have ${\sigma(i)=v_i}$ because in this case no other edge than $v_iv_i$ arrives at $v_i$ and appears in $M$. If not, let $k=\max I_v$. Since  $v'_k\not\in A$ and since in $M$ there is no edge arriving at $v'_k$ other that $v'_kv'_k$ and no edge starting from $v'_k$ other than $v'_kv_i$, then the only walk that can contain edge $v'_kv'_k$ is a walk starting from $v'_k$, arriving at $v$ and it must be of length $t_k$ to exhaust the power of $X_{v'_kv'_k}$ in $M$. Therefore, we must have ${\sigma'(k)=v'_k}$. Continuing with the same reasoning we show that $\sigma$ and $\sigma'$ are equal on $I_v$ for all $v$, which means ${\sigma'=\sigma}$. This shows that ${\det(N_\omega)\neq 0}$.

The degree of $N_\omega$ is clearly at most $n^2$. By the Schwartz-Zippel Lemma \cite[Theorem 7.1.4]{MP13}, there are at most $n^2q^{n^2 - 1}$ choices for the values of $X_e$ for which $\det(N_\omega) = 0$. Thus, there are at most ${n^2{n^2\choose n}q^{n^2 - 1}}$ choices for the values of $X_e$ for which some observation $\tau_\omega$ fails to be injective (recall that injectivity of $\tau_\omega$ is preserved by permutation of $\omega$). Since ${q > n^2{n^2\choose n}}$, we have ${q^{n^2} > n^2{n^2\choose n}q^{n^2 - 1}}$, and hence there exists a choice of values for all the variables $X_e$ such that $M$ is super-expansive.
\end{proof}

\newcommand\arr{\mathcal{A}}
\newcommand\cod{\mathcal{C}}
As an application, we shall see that any super-expansive (linear) network naturally gives rise to a (linear) orthogonal array and a maximum distance separable code. This can be formalized as follows (see \cite{HSS99,MS77} for an overview of the topic).

 An \emph{orthogonal array} of strength $s$ over alphabet $\(q\)$ and of index $1$ is a ${N\times M}$ array $\arr$ of elements of $\(q\)$ with ${s\leq N}$ and such that for any set of $s$ columns of $\arr$ no $s$-tuple appears two times. When $q$ is a prime power, we say the orthogonal array $\arr$ is linear if the set of rows is a vector space over $\GF(q)$. A code $\cod$ is a set of words from ${\(q\)^N}$ and its minimal distance ${d(\cod)}$ is the minimal Hamming distance between two distinct elements of $\cod$. When $q$ is a prime power, a code $\cod$ is linear if it forms a sub-vector space of ${\GF(q)^N}$. A maximum distance separable (MDS) code is a code verifying the equality in the so-called Singleton bound \cite{MS77}, \textit{i.e.} such that ${|\cod|=q^{N-d(\cod)+1}}$.

The link between these combinatorial objects and super-expansive networks is as follows. Given any $f\in\functions(n,q)$ let $\arr_f$ be the ${n^2\times q^n}$ array whose set of rows is ${\{L_x : x\in\(q\)^n\}}$ where 
\[L_x = (f(x)_1,f(x)_2,\ldots,f(x)_n,f^2(x)_1,\ldots,f^2(x)_n,\ldots,f^n(x)_1,\ldots,f^n(x)_n).\]

\begin{proposition} \label{prop:supermds}
  If ${f\in\functions(n,q)}$ is super-expansive then ${\arr_f}$ is an orthogonal array of strength $n$ and index $1$, and its set of rows is an MDS code of minimum distance ${n^2-n+1}$. If moreover $q$ is a prime power and $f$ is linear, then $\arr_f$ is linear and its set of rows is an MDS linear code.
\end{proposition}

\begin{proof}
  To any set of $n$ columns of $\arr_f$ is naturally associated ${\omega=\bigl(v_1,t_1),\ldots,(v_n,t_n)}$. The fact that $\tau_\omega$ is injective (by super-expansivity of $f$) exactly means that no pair of distinct lines $L_x$ can coincide on this set of columns. Hence, $\arr_f$ is an orthogonal array of strength $n$.  The fact that it also correspond to a MDS code is well-known and general \cite[Theorem 4.21]{HSS99}. Finally, it is straightforward to see that when $f$ is linear then $\arr_f$ is a linear and the corresponding code also.
\end{proof}

Using the classical bound of K. A. Bush on MDS codes \cite{B52}, we get a lower bound on the alphabet of a super-expansive network.

\begin{corollary} \label{cor:superbound}
  There is no super-expansive network with $n$ nodes over the alphabet $\(q\)$ if ${q\leq n^2-n}$.
\end{corollary}

\begin{proof}
  This follows immediately from the Bush bound \cite{B52}, which states that any orthogonal array of index $1$, strength $t$ over alphabet $\(q\)$ of size ${N\times M}$ verifies: ${N\leq t+q-1}$. The corollary follows from the previous proposition with ${N=n^2}$ and ${t=n}$.
\end{proof}

As a final remark, note that in the linear case, where $f(x) = xM$, then the corresponding MDS code has the following generator matrix:
\[
	G = ( M | M^2 | \dots | M^{n-1} ).
\]

\section{Links with expansivity in cellular automata}
\label{app:expca}

A topological dynamical system \cite{kurkabook} is a pair ${(F,X)}$ where $X$ is a compact metric space with distance $d$ and $F$ a continuous map. $F$ is said expansive if there is a real constant $\epsilon>0$ such that:
\[\forall x,y\in X, x\neq y\Rightarrow \exists t : d(F^t(x),F^t(y)\geq\epsilon.\]
In the general case the time $t$ is a positive integer and we speak about positive expansivity. If $F$ is bijective, then $t$ might be chosen negative (this is the original setting in \cite{U50}).

A (one-dimensional) cellular automaton is a topological dynamical system ${(F,Q^\Z)}$ where $Q$ is a finite alphabet and $F$ is defined through a local rule ${f: Q^V\rightarrow Q}$ with $V=[-r,\ldots,r]$ called neighborhood as follows: 
\[\forall x\in Q^\Z,\forall z\in\Z, F(x)_z = f(x_{|z+V})\]
where ${x_{|z+V}}$ denotes the map: ${i\in V\mapsto x_{z+i}}$. $F$ is positively expansive if and only if the following trace map is bijective \cite[Proposition 5.48]{kurkabook}: 
\[x \mapsto (x_{|V},F(x)_{|V},F^2(x)_{|V},\ldots)\]
or equivalently if and only if
\[\forall x,y\in X, x\neq y\Rightarrow \exists t : F^t(x)_{|V}\neq F^t(y)_{|V}.\]

Expansivity (positive or not) in cellular automata has received a lot of attention \cite{blanchardmaass,Pivato11,nasu06} and is still an active direction of research \cite{JalonenK18}, one of the main open problem being the decidability of the property (see \cite[Problem 19]{BoyleOpen} or \cite[Problem 7]{Kari05}). 

Expansive cellular automata and expansive automata networks can be linked in two ways:
\begin{enumerate}
\item we can see a cellular automaton $F$ as an automata network on the infinite graph ${(\Z,E)}$ where ${(i,j)\in E}$ if and only if ${|i-j|\leq r}$ (taking the notation above). This graph is always strong and coverable (because it contains self-loops) and $F$ as a cellular automaton is positively expansive if and only if it is quasi-expansive as an (infinite) automata network (see Theorem~\ref{th:expansive} and previous definitions).
\item we can also restrict a cellular automaton $F$ to periodic configurations of period $n$. In this case it can be seen as a standard automata network $F_n$ on the finite graph ${(\Z/n\Z,E)}$ where ${(i,j)\in E}$ if and only if ${|i-j|\leq r}$. If $F$ as a cellular automaton is positively expansive, then for any $n$, the automata network $F_n$ is quasi-expansive. The converse is false as the shift cellular automaton 
  ${F(x)_z = x_{z+1}}$  is not positively expansive while all its restrictions $F_n$ are quasi-expansive.
\end{enumerate}

\end{document}